\documentclass[12pt]{iopart}
\usepackage{graphicx}

\renewcommand{\Re}{\mathop{\mathrm{Re}}\nolimits}
\renewcommand{\Im}{\mathop{\mathrm{Im}}\nolimits}
\newcommand{\D}{\mbox{\rm d}}

\begin{document}

\title[Nonclassicality of noisy quantum states]{Nonclassicality of
noisy quantum states} 
\author{
A A Semenov\dag, D Yu Vasylyev\dag\ddag, and B I Lev\dag}
\address{\dag  Institute of Physics,
National Academy of Sciences of Ukraine, 46 Prospect Nauky,
UA-03028 Kiev, Ukraine}
\address{\ddag  Fachbereich Physik, Universit\"{a}t Rostock,
Universit\"{a}tsplatz 3, D-18051 Rostock, Germany}
\eads{\mailto{sem@iop.kiev.ua}, \mailto{dim@iop.kiev.ua},
\mailto{lev@iop.kiev.ua}}

\begin{abstract}
Nonclassicality conditions for an oscillator-like system
interacting with a hot thermal bath are considered. Nonclassical
properties of quantum states can be conserved up to a certain
temperature threshold only. In this case, affection of the thermal
noise can be compensated via transformation of an observable,
which tests the nonclassicality (witness function). Possibilities
for experimental implementations based on unbalanced homodyning
are discussed. At the same time, we demonstrate that the scheme
based on balanced homodyning cannot be improved for noisy states
with proposed technique and should be applied directly.
\end{abstract}

  \maketitle

\section{Introduction}

Nonclassical properties play a crucial role in understanding
fundamentals of quantum physics. This concept is the main
theoretical background for many applications, including quantum
information processing \cite{QI}. Usually, the nonclassicality
manifests itself in specific properties of quantum statistics,
which sometimes cannot be described in the framework of the
probability theory \cite{Klyshko}.

This phenomenon was firstly considered in the famous work by
Einstein, Podolsky, and Rosen \cite{EPR} for demonstartion of
contradictions between quantum mechanics and the concept of
``local realism''. As it was shown by Bell \cite{Bell}, the
later leads to some inequalities violated for usual quantum
mechanics. Experiments of Aspect with collaborators \cite{Aspect}
confirmed this fact as well as assumption that quantum phenomena
are characterized by a specific feature, nonlocality, which cannot
be explained in classical terms. Note that nowadays there exist
some other criteria for testing the presence of this kind of
nonclassicality (entanglement) \cite{Peres, Horodecki, Simon,
Shchukin}.

Reconstruction of the Wigner probability distribution
\cite{Wigner} in experiments with quantum tomography \cite{VogelK,
Smithey, WelschRev} demonstrates appearance of its negative
values. This peculiarity cannot be explained in the framework of the
probability theory, hence it does not have any classical
counterparts. However, quadrature squeezing \cite{Slusher} as well
as sub-Poissonian statistics of photon counts \cite{Short}, being
well-known examples of nonclassicality, is still possible for
some states with completely positive Wigner function. This is
explained by the fact that these features correspond to negative
values of dispersions for some normally ordered observables, which
are naturally determined via the Glauber-Sudarshan $P$-function
\cite{Glauber, Sudarshan}. Properties of this distribution differ
from those for the Wigner function. For example, there exist states
characterized by both non-negative Wigner function and
non-positive Glauber-Sudarshan $P$-function.

Therefore, the class of states characterized by non-positive
$P$-function includes the states characterized by non-positive
Wigner function, sub-Poissonian statistics of photon counts and
quadrature squeezing. Hence, following to the works \cite{Vogel,
Richter}, we will consider the nonclassicality as non-positivity
of the $P$-function. Unlike the case of the Wigner function, this
definition cannot be applied directly because of a very strong
singularity of the $P$-function for nonclassical states.

The Bochner criterion \cite{Bochner} allows one to formulate
observable conditions for experimental testing the nonclassicality
\cite{Richter}. Arguing in this way, we conclude that the
$P$-function is a positive-definite one (or can be interpreted as
probability density) if and only if for any function $f(\alpha)$
the following inequality is satisfied:
\begin{equation}
\int\limits_{-\infty}^{+\infty}{\D^2{\alpha}}\int\limits_{-\infty}^{+\infty}{\D^2\beta}\
\Phi(\alpha-\beta)f(\alpha)f^{\ast} (\beta) \ge0, \label{2}
\end{equation}
where $\Phi(\beta)$ is the characteristic function of the
$P$-distribution, {\em i.e.} its Fourier transform
\begin{equation}
\Phi (\beta) = \int\limits_{-\infty}^{+\infty}{\D^2\alpha\
P(\alpha)\exp(\alpha^{\ast}\beta-\alpha \beta^{\ast}}).\label{1}
\end{equation}

Hence, in order to test quantum states on the nonclassicality, it
is sufficient to find such a function $f(\alpha)$ that violates
the inequality (\ref{2}). An important example (as a matter of
fact it was used in \cite{Richter}) is the discrete variant of the
Bochner criterion when this function is taken in the following
form:
\begin{equation}
f(\alpha)=\sum_{k}\xi_{k}\
\delta(\alpha-\alpha_{k}),\label{discrete}
\end{equation}
where $\xi_{k}$, $\alpha_{k}$ are some arbitrary complex numbers.
The experimental implementation of this criterion for optical
fields was described in \cite{Lvovsky}.

The inequality (\ref{2}) can be rewritten in another equivalent
form. For this purpose we introduce the object
$\mathcal{W}(\alpha)$, which following \cite{Korbicz} we will
refer to as {\em the witness function} and define as following
\begin{equation}
\mathcal{W}(\alpha)=|{g(\alpha)}|^{2}\ge0,\label{4}
\end{equation}
where $g(\alpha)$ is the Fourier image of $f(\alpha)$. It is easy
to see that the inequality (\ref{2}) takes now the following form:
\begin{equation}
\int\limits_{-\infty}^{+\infty}{\D^{2}{\alpha}P(\alpha
)\mathcal{W}(\alpha)}\ge0 \label{3}.
\end{equation}
In other words,  expression (\ref{3}) means that the mean value
of some operator $\hat \mathcal{W}$ must be greater or equal to
zero.  This operator is defined in such a way that its
normally ordered symbol is the witness function
$\mathcal{W}\left(\alpha\right)$, {\em i.e.}
\begin{equation}
\mathcal{W}\left(\alpha\right)=\langle \alpha | \hat \mathcal{W} |
\alpha\rangle, \label{woperator}
\end{equation}
where $|\ \alpha\rangle $ is coherent state. Hence, if we succeed
to find the operator $\hat \mathcal{W}$ satisfying
(\ref{woperator}) and (\ref{4}), such that its mean value is less
than zero (or  condition (\ref{3}) fails to obey), then we can
assert that the nonclassicality is inherent for the given state.
As it was noted in \cite{Korbicz}, the concept of the witness
function can be used for testing other kinds of nonclassicality as
well.

The main problem for testing the nonclassicality for realistic
systems is the decoherence \cite{Joos}. It is reasoned by
uncontrolled interaction of the system with an environment that
leads to substantial effects on nonclassical properties. Depending
on the system, the environment may have various physical nature. One
of the well-known examples of such a system is an internal mode of
high-$Q$ cavity \cite{Collet, Gardzoller}. In this case, the
number of external modes plays a role of the environment, which
interacts with the system through the semitransparent mirror.
Another example is absorption and scattering of the
electromagnetic field by cavity mirrors while intracavity mode is
extracted outside for the further use \cite{Khanbekyan}.
Absorption and scattering processes also take place while
transferring a quantum electromagnetic signal through a
semitransparent plate \cite{Chizhov}, waveguide {\em etc}.

The number of thermal photons in the optical domain of the
electromagnetic radiation is negligibly small. Hence, in this case
the environment can be regarded as being in the vacuum state. The
main problem is that the modern technologies in many cases do not
afford to produce the optical devices with small interaction of
the electromagnetic field and the absorbing medium. Especially
this is apparent for optical high-$Q$ cavities, where resulting
outgoing pulse includes just near 50\% of the initial intracavity
mode \cite{Khanbekyan}. Somewhat the microwave cavities are devoid
of this shortcoming. In this case, the constant of interaction
between field and absorption system is comparatively small (see the
discussion in \cite{Khanbekyan}). However, the microwave domain is
characterized by the presence of great number of thermal photons.
This causes more serious difficulties in testing nonclassical
properties of quantum states. Thus, there arises a natural
question about a balance between the constant of interaction and
temperature of environment for optimal detection of the
nonclassicality. This is the subject of the presented paper.

The paper is organized as follows. In Sec.~\ref{S2}, we obtain an
equation, which describes the evolution of the Glauber-Sudarshan
$P$-distribution of an oscillator-like system under the thermal
noise influence. In Sec.~\ref{S3}, we consider how to redefine the
witness function and optimize testing the nonclassicality for
noisy state. Application of this method with the schemes of
unbalanced homodyne detection is considered in Sec.~\ref{S4}. In
Sec.~\ref{S5}, we demonstrate that the discrete form of the Bochner
criterion does not allow a further improvement and its presented
form is the best choice even for the noisy states. An example of
single-photon Fock number state is presented in Sec.~\ref{S6}. The
last section contains the summary and conclusion.

\section{Quantum state of system under the noise influence}
\label{S2}

We will describe open quantum systems using the input-output
formalism, considered in the book \cite{Gardzoller}. It is worth
noting that we focus on a class of linear systems only. It gives
us a possibility of considering a wide enough class of experiments
with quantum electromagnetic field of low intensity.

\begin{figure}[ht]
\begin{center}
\includegraphics[clip=]{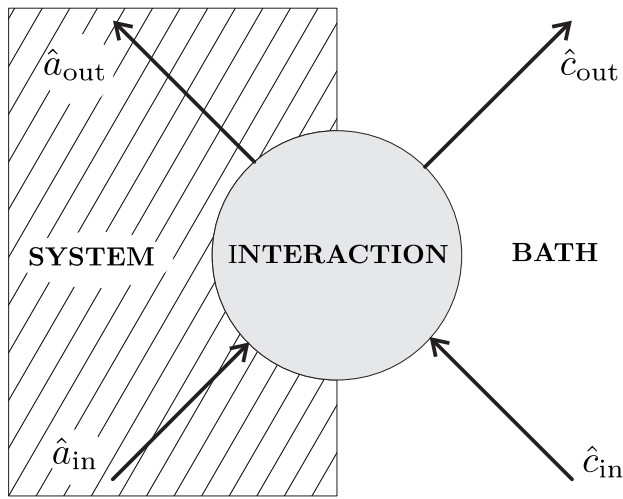}
\end{center}
\caption{\label{fig1} The model of open quantum system in terms of
the input-output formalism. $\hat{a}_\mathrm{in}$ and
$\hat{c}_\mathrm{in}$ are operators of the system and environment
before interaction, $\hat{a}_\mathrm{out}$ and
$\hat{c}_\mathrm{out}$ correspond to these operators after
interaction.}
\end{figure}

Following the input-output formalism, let us consider an open
quantum system as a device with two input-output ports. One of
them corresponds to a system  and another one corresponds to
the bath (see Fig.~\ref{fig1}). Let the operator $\hat
a_\mathrm{in}$ describes a system before interaction. For example,
it can be an operator of the input-field mode. The operator $\hat
c_\mathrm{in}$ describes degrees of freedom of an environment
before interaction, {\em e.g.}, at the initial point of time. In
the same manner we will describe the system after interaction in
terms of the operator $\hat a_\mathrm{out}$, which can be
interpreted as the output-field mode operator. We also suppose
that these operators satisfy the usual bosonic commutation
relations
\begin{eqnarray}
\left[\hat a_\mathrm{in}, \hat a_\mathrm{in}^{\dag}\right]=1,\\
\left[\hat a_\mathrm{out}, \hat a_\mathrm{out}^{\dag}\right]=1,\\
\left[\hat c_\mathrm{in}, \hat c_\mathrm{in}^{\dag}\right]=1.
\end{eqnarray}
Without loss of generality, we can describe the evolution of the
system in terms of a linear input-output relation between these
operators
\begin{equation}
\hat a_\mathrm{out} = \sqrt{\eta}\ \hat a_\mathrm{in}+
\sqrt{1-\eta}\ \hat c_\mathrm{in},\label{aout1}
\end{equation}
where $0\le\eta\le1$ is efficiency of the input-output processes.

\begin{figure}[ht]
\begin{center}
\includegraphics[clip=]{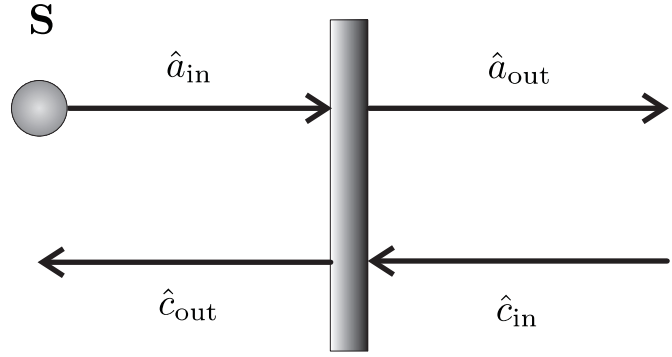}
\end{center}
\caption{\label{fig2} Transmission of a signal through a partially
transparent plate. $\hat a_\mathrm{in}$ and $\hat a_\mathrm{out}$
are input and output operators of the signal. $\hat c_\mathrm{in}$
and $\hat c_\mathrm{out}$ are operators of the field-mode passing
in the inverse direction. Latter plays a role of the bath
interacting with the signal (system) through a partially
transparent plate.}
\end{figure}

Let us regard some elementary examples. First of all consider
transmission of a signal through a partially transparent plate
(see Fig.~\ref{fig2}). A source of quantum electromagnetic field
$\mathrm{S}$ radiates the signal $\hat a_\mathrm{in}$ and, after
transmitting through the plate, the output mode $\hat a_\mathrm{out}$
is detected. The bath system in this case is the
electromagnetic-field modes ($\hat c_\mathrm{in}, \hat
c_\mathrm{out}$), which pass in another direction. The plate plays
a role of a device, which is responsible for a linear interaction
between  both the modes, {\em i.e.} between the system and the
environment. Hence, Eq.~(\ref{aout1}) is the well-known
input-output relation for the partially transparent plate to
within a phase multiplier(see {\em e.g.} \cite{VogelBook}). The
efficiency $\eta$ is connected with the transmission coefficient
$T$ of the plate in the following form:
\begin{equation*}
\eta=\left|\ T\right|^2.
\end{equation*}

Another example is the process of quantum-state extraction from a
high-$Q$ cavity \cite{Khanbekyan}. In this case, the operator
$\hat a_\mathrm{in}$ can be interpreted as the operator of an
intracavity mode at the initial time. The operator $\hat
a_\mathrm{out}$ corresponds to the non-monochromatic mode leaking
from the cavity. The processes of absorption and scattering by
mirrors can be considered as interaction between the system and
the bath. Hence, the operator $\hat c_\mathrm{in}$ corresponds to
the absorption system of the mirror and scattering modes of field.
Corresponding input-output relation is considered in work
\cite{Khanbekyan}. The efficiency of this process is closely
related to two components of the cavity decay rate:
$\gamma_\mathrm{rad}$ which is responsible for the output and
$\gamma_\mathrm{abs}$ which is responsible for the absorption and
scattering
\begin{equation}
\eta=\frac{\gamma_\mathrm{rad}}{\gamma_\mathrm{rad}+
\gamma_\mathrm{abs}}.
\end{equation}

As the next step we will rewrite the input-output relation
(\ref{aout1}) in the Schr\"odinger picture of motion, {\em i.e.}
consider transformation of the density operator under the noise
influence. As it was  already mentioned in the introduction, it is
convenient to describe the nonclassicality using the Glauber-Sudarshan
$P$-representation \cite{Glauber, Sudarshan}. The characteristic
function (\ref{1}) of the output field can be written down as
follows:
\begin{equation}
\Phi_\mathrm{out}(\beta)= \Tr\left\{\ \hat\rho\ e^{\hat
a_\mathrm{out}^{\dag}\beta-\ \hat a_\mathrm{out}\beta^{\ast}
+\frac{|\beta|^2}{2}} \right\}, \label{char1}
\end{equation}
where $\hat\rho$ is density operator of the system and the bath.
Moreover, we suppose that at the initial time it can be decomposed
as
\begin{equation}
\hat\rho =\hat\rho_\mathrm{in}\otimes
\hat\rho_\mathrm{bath}.\label{rho}
\end{equation}

Substituting the input-output relations (\ref{aout1}) into
Eq.~(\ref{char1}) and taking into account (\ref{rho}), one can
obtain the
following expression for the characteristic function of the output state: 
\begin{equation}
\Phi_\mathrm{out}(\beta)=\Phi_\mathrm{in}(\beta\sqrt{\eta})
\Phi_\mathrm{bath}(\beta\sqrt{1-\eta}),
\end{equation}
where $\Phi_\mathrm{in}(\beta)$ is the characteristic function of
the input (noiseless) state of the system and
$\Phi_\mathrm{bath}(\beta)$ is the characteristic function of the
input state of the bath. We suppose that initially the bath is in
the thermal state with temperature $T$ and the mean value of
thermal photons $\bar n=\left[\exp\
(\frac{\hbar\omega}{kT})-1\right]^{-1}$, {\em i. e.},
\begin{equation}
\hat \rho_\mathrm{bath}= \frac{1}{1+\bar n} \left(\frac{\bar
n}{1+\bar n}\right)^{\hat n},
\end{equation}
\begin{equation}
\Phi_\mathrm{bath}(\beta)=e^{-|\beta|^2 \bar{n}}.
\end{equation}
Hence, the characteristic function of the output (noisy) state is
written in the following form
\begin{equation}
\Phi_\mathrm{out}(\beta)= \Phi_\mathrm{in}(\beta{\sqrt{\eta}})\
e^{-|\beta|^2 \bar{n}(1-\eta)}.\label{A}
\end{equation}
One can easily see that for
\begin{equation}
\bar{n}=\frac{\eta}{1-\eta} \label{B}
\end{equation}
the characteristic function (\ref{A}) turns into a characteristic
function for the positive-definite Husimi-Kano $Q$-distribution
\cite{Husimi, Kano} of a certain state. Thus, for such values of
$\bar n$ the $P$-function is always positive. One can say the same
if the number of thermal photons is greater than the value
determined by  equation (\ref{B}). Therefore, Eq.~(\ref{B})
defines {\em thermal threshold of the nonclassicality}. In other
words, if the number of thermal photons in the bath is greater
than the value given by Eq. (\ref{B}), the nonclassicality in the
sense of negative values of $P$-function always vanishes. However
only the fact that the number of thermal photons is less than the
thermal threshold defined by Eq.~(\ref{B}), cannot be considered
as a sufficient condition of the nonclassicality.

Using the inverse Fourier transform we get from Eq.~(\ref{A}) the
relation for the $P$-function of the state in the following form:
\begin{equation}
P_\mathrm{out}(\alpha)=\frac{1}{\eta} \exp\left [{\ \bar n
(1-\eta)\ \Delta_\alpha}\right ]
P_\mathrm{in}\left(\frac{\alpha}{\sqrt{\eta}}\right ),\label{Peq}
\end{equation}
where $\Delta_\alpha=
\frac{\partial^2}{\partial\left(\Re\alpha\right)^2}+
\frac{\partial^2}{\partial\left(\Im\alpha\right)^2}$. This means
that the $P$-function for the output state satisfies the
diffusion-like differential equation
\begin{equation}
\frac{\partial}{\partial \bar n}P_\mathrm{out}(\alpha, \bar
n)=\left( 1-\eta \right)\Delta_\alpha P_\mathrm{out}(\alpha, \bar
n )\label{Peq2}
\end{equation}
with the ``initial'' condition
\begin{equation}
P_\mathrm{out}(\alpha, 0)=\frac{1}{\eta}\ P_\mathrm{in}\left (
\frac{\alpha}{\sqrt\eta}\right ).
\end{equation}

Therefore, taking into account above mentioned  we can conclude
that under the thermal noise influence the $P$-function of the
system transforms according to the diffusion-like differential
equation (\ref{Peq2}), where the mean number of the thermal
photons plays the role of ``time variable''. This means that with
growing the number of thermal photons in the environment, $P$-function
of the system is smoothed. For a certain value of $\bar{n}$, which is
less or equal to the thermal threshold (\ref{B}), domains of its
negative values disappear. In \cite{Lee} this critical number of thermal photons 
(in our notations this is $\frac{1-\eta}{\eta}\bar n$) has been proposed to be used as a measure of nonclassicality.

\section{Witness function for the output state}
\label{S3}

Let us suppose that a quantum state of the system already
manifests the nonclassicality. In other words, according to
Eq.~(\ref{3}), there exists such a witness function
$\mathcal{W}(\alpha)$ that
\begin{equation}
\bar \mathcal{W}=\int\limits_{-\infty}^{+\infty} \D^{2}\alpha\
P_\mathrm{in}(\alpha)\ \mathcal{W}(\alpha)<0.\label{C}
\end{equation}
Consider corresponding state of the output mode given by
Eq.~(\ref{Peq}). To order it manifests the property of the
nonclassicality as well, another witness function
$\mathcal{W}_\mathrm{th}(\alpha,\bar n)$ should exist and the
following inequality should be true:
\begin{equation}
\bar{\mathcal{W}}_\mathrm{th}=\int\limits_{-\infty}^{+\infty}
\D^{2}\alpha\ P_\mathrm{out}(\alpha)\
\mathcal{W}_\mathrm{th}(\alpha,\bar n )<0.\label{D}
\end{equation}

We now want to check whether $\mathcal{W}_\mathrm{th}(\alpha,\bar
n)$ can be chosen in such a form that $\bar
\mathcal{W}=\bar{\mathcal{W}}_\mathrm{th}<0$. For doing this let
us substitute Eq.~(\ref{Peq}) for the $P$-function of the output
state into Eq.~(\ref{D})
\begin{equation}
\bar{\mathcal{W}}_\mathrm{th}=\int\limits_{-\infty}^{+\infty}
\D^2\alpha\ \left[\frac{1}{\eta}\ \exp\left [\bar n(1-\eta)\
\Delta_\alpha
\right]P_\mathrm{in}\left(\frac{\alpha}{\sqrt{\eta}}\right)\right]\
\mathcal{W}_\mathrm{th}(\alpha,\bar n) .\label{E}
\end{equation}
Taking into account that the diffusion operator is a Hermitian one
and penetrating the change of variables in the last equation we
can rewrite it as follows:
 \begin{equation}
\bar \mathcal{W}_\mathrm{th}=\int\limits_{-\infty}^{+\infty}
\D^2\alpha\ P_\mathrm{in}(\alpha)\left[\ \exp\left (\frac{\bar
n(1-\eta)}{\eta}\ \Delta_\alpha \right)\
\mathcal{W}_\mathrm{th}(\alpha\sqrt{\eta},\bar n)\right
].\label{F}
\end{equation}
Comparing Eqs.~(\ref{C}), (\ref{F}), one can conclude that $\bar
\mathcal{W}=\bar{\mathcal{W}}_\mathrm{th}$ if
\begin{equation}
 \mathcal{W}(\alpha)= \exp\left [\frac{\bar
n(1-\eta)}{\eta}\ \Delta_\alpha \right]\
\mathcal{W}_\mathrm{th}(\alpha\sqrt{\eta},\bar n).\label{G}
\end{equation}
Inverting this equation, one obtains the following expression for
the witness function, which tests the nonclassicality for the
output (noisy) state
\begin{equation}
\mathcal{W}_\mathrm{th}(\alpha,\bar n)=\exp\left[-\bar n(1-\eta)\
\Delta_\alpha\right] \
\mathcal{W}\left(\frac{\alpha}{\sqrt{\eta}}\right).\label{H}
\end{equation}
This means that we can find the new witness function as a solution
of diffusion-like differential equation with a negative diffusion
coefficient
\begin{equation}
\frac{\partial}{\partial \bar n}\mathcal{W}_\mathrm{th}(\alpha,
\bar n)=-(1-\eta)\Delta_\alpha\mathcal{W}_\mathrm{th}(\alpha, \bar
n),\label{H1}
\end{equation}
and the following ``initial'' condition
\begin{equation}
 \mathcal{W}_\mathrm{th}(\alpha,0)=\mathcal{W}\left (\frac{\alpha}{\sqrt{\eta}}\right).\label{H2}
\end{equation}

Therefore, for a noisy state the witness function can be redefined
in such a way that testing the nonclassicality gives a result
equal to the noiseless case. The fact that Eq.~(\ref{H1}) is a
diffusion-like equation with negative diffusion coefficient means
that with growing the number of thermal photons $\bar{n}$ one has
to choose a sharper witness function for obtaining the same
result. It is clear that this possibility exists up to a certain
temperature threshold only, which however can be less than the
value defined by Eq.~(\ref{B}) but cannot be greater. Beyond it
the solution of Eq.~(\ref{H1}) may have both strong singularities
and domains of negative values.

\section{Testing the nonclassicality in experiments with unbalanced
homodyning} \label{S4}

Unbalanced homodyning, proposed in \cite{Wallentowitz}, allows one
to test the nonclassicality with an important class of witness
functions, which have a form of the Gauss distribution
\begin{equation}
\mathcal{W}(\alpha)=\frac{1}{\pi a^2}\
\exp\left[-\frac{|\alpha-\gamma|^2}{a^2}\right].\label{J}
\end{equation}
Firstly consider the case of a noiseless quantum state, {\em i.e.}
the input field. Inserting the witness function (\ref{J}) into
Eq.~(\ref{C}) we get that $\bar \mathcal{W}$ can be regarded as
value of some $s$-parameterized distribution \cite{Glauber2} in
point $\gamma$
\begin{equation}
\bar \mathcal{W}=\frac{1}{\pi a^2}\int\limits_{-\infty}^{+\infty}
\D^2\alpha\ P_\mathrm{in}(\alpha)\
\exp\left[-\frac{|\alpha-\gamma|^2}{a^2}\right]=P_\mathrm{in}
(\gamma,s),\label{L}
\end{equation}
where
\begin{equation}
s=1-2a^2.\label{s}
\end{equation}
It is clear that it never tests the nonclassicality for $s\leq
-1$, {\em i.e.} $a^2\geq 1$.

\begin{figure}[ht]
\begin{center}
\includegraphics[clip=]{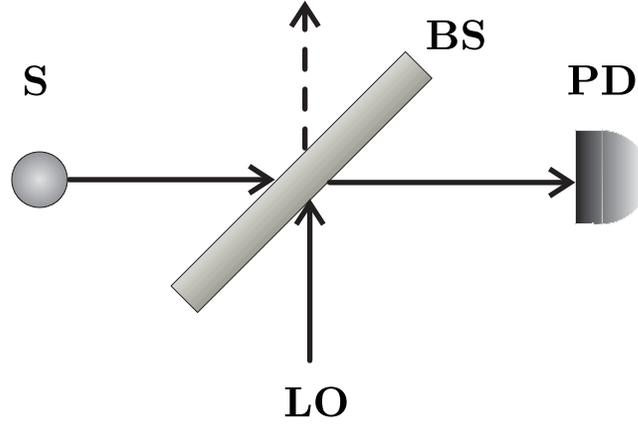}
\end{center}
\caption{\label{fig3} The scheme of unbalanced homodyne
detection.}
\end{figure}

The scheme of the corresponding experiment is presented in
Fig.~\ref{fig3}. The signal, which is tested for the
nonclassicality, is combined through a beam splitter with the
local-oscillator field. The photon-counting detector, placed in an
output port, is used for the measurement of counting distributions
$P_{n}^\mathrm{in}(\gamma, \eta_\mathrm{h})$. They depend on the
coherent amplitude $\gamma$, and the overall efficiency of the
homodyning $\eta_\mathrm{h}$, which are expressed in terms of
amplitude of the local oscillator $\gamma_\mathrm{lo}$,
transmission $T$ and reflection coefficients $R$  of the
beam splitter and the efficiency of photon counting $\zeta$
\begin{eqnarray}
\gamma=-\frac{R}{T}\gamma_\mathrm{lo},\\
\eta_\mathrm{h}=\zeta\left|T\right|^2.
\end{eqnarray}
As it was shown in \cite{Wallentowitz} the value of $\bar
\mathcal{W}$, given by Eq.~(\ref{L}), can be reconstructed from
the probabilities of photon counts  using the following
expression
\begin{equation}
\bar \mathcal{W}=\frac{1}{\pi
a^2}\sum\limits_{n=0}^{+\infty}\left[-\frac{1-\eta_\mathrm{h} a^2
}{\eta_\mathrm{h} a^2 }\right]^n P_{n}^\mathrm{in}(\gamma,
\eta_\mathrm{h}).\label{Wall}
\end{equation}

Assume that the nonclassicality can be detected for the noiseless
signal with the above described  procedure. This gives us a
possibility to find such a witness function, which tests the
nonclassicality for the corresponding noisy state. It can be
obtained by resolving the diffusion-like equation (\ref{H1}), with
`initial' condition (\ref{H2}) specified by  function
(\ref{J}). The solution is written  as follows:
\begin{equation}
\mathcal{W}_\mathrm{th}(\alpha)=\frac{1}{\pi \left(a^2-\bar
n\frac{\eta}{1-\eta}\right)}\
\exp\left[-\frac{\left|\frac{\alpha}{\sqrt{\eta}}-\gamma\right|^2}{a^2-\bar
n\frac{\eta}{1-\eta}}\right].\label{K}
\end{equation}

It is worth noting that this solution is defined just for
\begin{equation}
\bar n\leq a^2\frac{\eta}{1-\eta}.\label{SCondition}
\end{equation}
For other values of $\bar n$, it is not positive definite and has
strong singularities. This is a typical property for the solution
of diffusion equation with negative diffusion coefficient. The
last expression defines the thermal threshold for the scheme of
unbalanced homodyning. Taking into account that the maximal value
of $a^2$ is $1$, one immediately obtains Eq.~(\ref{B}).

Corresponding value of $\bar{\mathcal{W}}_\mathrm{th}$ as well as
$s$-parameterized distribution for the noiseless signal can be
reconstructed from the probabilities of photon counts for noisy
signal $P_{n}^\mathrm{out}(\gamma, \eta_\mathrm{h})$  using
the following expression:
\begin{eqnarray}
&&\bar {\mathcal{W}}_\mathrm{th}=P_\mathrm{in} (\gamma,s)\nonumber\\
&&=\frac{\eta}{\pi \left(a^2-\bar
n\frac{\eta}{1-\eta}\right)}\sum\limits_{n=0}^{+\infty}
\left[-\frac{1-\eta_\mathrm{h} \left(a^2-\bar
n\frac{\eta}{1-\eta}\right) }{\eta_\mathrm{h} \left(a^2-\bar
n\frac{\eta}{1-\eta}\right)}\right]^n
P_{n}^\mathrm{out}(\gamma\sqrt{\eta},
\eta_\mathrm{h}),\label{Modified}
\end{eqnarray}
where $s$ is defined by Eq.~(\ref{s}). A disadvantage of this
method is a fact that for some quantum states the series defined
by Eqs.~(\ref{Wall}), (\ref{Modified}) may diverge (see
\cite{Wallentowitz}). This gives some restrictions for the application
of this method.

\section{Discrete form of the Bochner criterion}
\label{S5}

Discrete form of the Bochner criterion allows one to formulate
observable conditions \cite{Richter} based on the technique of
balanced homodyning. Combining Eqs.~(\ref{2}) and
(\ref{discrete}), one obtains the following inequality:
\begin{equation}
\sum\limits_{k,l}\Phi_\mathrm{in}\left(\alpha_k-\alpha_l\right)\xi_{k}
\xi_{l}^{\ast}\geq 0. \label{VogelCond}
\end{equation}
In case if it violates for some values of parameters ({\em i.e.},
the quadratic form is not a positive-definite one), the state has
nonclassical properties. In other words, the nonclassicality
appears if at least one minor of the matrix
$\Phi_{l,k}^\mathrm{in}=\Phi_\mathrm{in}\left(\alpha_k-\alpha_l\right)$
has a negative value for certain values of $\alpha_k$. This matrix
depends just on the values of characteristic function for the
$P$-distribution in certain points. The latter can be reconstructed
via measurements of quadratures. For the electromagnetic fields,
this can be performed with procedure of the balanced homodyne
detection \cite{Yuen}. Corresponding methods are well known for
other systems \cite{WelschRev}.

The witness function $\mathcal{W}(\alpha)$ for the discrete form
of the Bochner criterion can be easily found utilizing the
definition given by Eq.~(\ref{4})
\begin{equation}
\mathcal{W}(\alpha)=\sum\limits_{k,l}\xi_{k}\xi_{l}^{\ast}\
\exp\left[\alpha^{\ast}(\alpha_k-\alpha_l)-\alpha(\alpha_k^{\ast}
-\alpha_l^{\ast})\right].\label{DiscreteWitness}
\end{equation}
Now we may consider a question whether it is possible to find the
witness function $\mathcal{W}_\mathrm{th}(\alpha)$ which tests
the nonclassicality for a noisy state in the same manner as for
the corresponding noiseless one. For this purpose, we should
resolve the diffusion-like equation (\ref{H1}), with ``initial''
condition (\ref{H2}) specified by the function
(\ref{DiscreteWitness}). The result is written  in the
following form:
\begin{equation}
\mathcal{W}_\mathrm{th}(\alpha)=\sum\limits_{k,l}\xi_{k}\xi_{l}^{\ast}\
e^{\bar{n}\frac{1-\eta}{\eta}\left|\alpha_k-\alpha_l\right|^2}
\exp\left[\alpha^{\ast}\frac{\alpha_k-\alpha_l}{\sqrt{\eta}}
-\alpha\frac{\alpha_k^{\ast}-\alpha_l^{\ast}}{\sqrt{\eta}}\right].
\label{DiscreteWitnessTh}
\end{equation}
This function is not a positive-definite one and consequently it
can not be considered as a witness function. Therefore, the witness
function in the form of Eq.~(\ref{DiscreteWitness}) cannot be
improved with described method. Hence, the nonclassicality
conditions in form of the discrete Bochner criterion
\cite{Richter} should be applied directly which is the best choice
even for the noisy state.

\section{An example: Fock state}
\label{S6}

Single-photon Fock number state with density operator
\begin{equation}
\hat\rho_\mathrm{in}=| 1 \rangle\langle 1 |.\label{PureFock}
\end{equation}
is a good candidate for the experimental realization of the proposed
method. Generation of this state with using the frequency
down-conversion process and testing it for the nonclassicality
with the application of the balanced homdyne detection is reported in
\cite{Lvovsky}. $s$-parameterized distribution for the
single-photon Fock state has the following form:
\begin{equation}
P_\mathrm{in}\left(\alpha,s\right)= \left\{
\begin{array}{ll}
\frac{2}{\pi(1-s)^3}\left(4\left|\alpha\right|^2-1+s^2\right)\exp
\left[-\frac{2|\alpha|^2}{1-s}\right], &-1\leq s< 1\\
\left(1+\frac{1}{4}\partial^{2}_{\alpha^{\ast}
\alpha}\right)\delta(\alpha),& s=1
\end{array}
\right. ,
\end{equation}
This is a regular function for $-1\leq s< 1$, and a distribution
with very strong singularity for $s=1$, {\em i.e.} for the
Glauber-Sudarshan $P$-function. It is clear that this state has
nonclassical properties and non-positive-definite phase-space
distributions for all values of the parameter $s\neq -1$.

Different losses in experimental set-up lead to admixing the vacuum
state into the density operator (\ref{PureFock}). Hence, the
resulting state has the form of the following statistical mixture:
\begin{equation}
\hat\rho_\mathrm{out}=\eta| 1 \rangle\langle 1 |+
\left(1-\eta\right)| 0 \rangle\langle 0 |.\label{AddFock}
\end{equation}
This is a result of interaction between the field mode and
zero-temperature bath (where the mean number of thermal photons
$\bar n$ is negligible). As it was shown in \cite{Lvovsky}, the
nonclassicality can be tested, at least in principle, for any
value of the efficiency $\eta$. Such situation is usual for the
optical domain. We consider a more general case, when the bath
does have non-zero temperature, that is typical for the microwave
domain, vibrational motion of trapped atom, {\em etc}.

The Glauber-Sudarshan $P$-function for the thermal noisy state can
be obtained from Eq.~(\ref{Peq}) and is written  as follows:
\begin{equation}
P_\mathrm{out}(\alpha)=\frac{1}{\eta}
P_\mathrm{in}\left(\frac{\alpha}{\sqrt{\eta}}, s^\prime\right
),\label{P1}
\end{equation}
where
\begin{equation}
s^\prime=1-2\bar n \frac{1-\eta}{\eta}.\label{sprime}
\end{equation}
The right-hand side of this equation in the case of single-photon Fock
state has non-positive values for $s^\prime>-1$. Therefore, taking
into account Eq.~(\ref{sprime}) we conclude that this state under
the thermal noise influence preserves nonclassical properties up
to the thermal threshold given by Eq.~(\ref{B}).

Application of the unbalanced homhodyning scheme, considered in
Sec.~\ref{S4}, means testing the nonclassicality with the witness
function $\mathcal{W}\left(\alpha\right)$ given by Eq.~(\ref{J}).
Hence, according to Eq.~(\ref{L}) for single-photon (noiseless)
Fock state one has the following value for the quantity
$\bar{\mathcal{W}}$:
\begin{equation}
\bar{\mathcal{W}}=P_\mathrm{in}\left(\gamma,
1-2a^2\right)=\frac{1}{\pi a^6}
\left(\left|\gamma\right|^2+a^2\left(a^2-1\right)\right)
\exp\left[-\frac{\left|\gamma\right|^2}{a^2}\right].\label{FockBarW}
\end{equation}
It has negative values for any $a^2<1$ and, moreover, shows
non-positivity of the phase-space distribution with $s=1-2a^2$.

Applying the same witness function for single-photon Fock state
under thermal noise influence, one obtains the following value:
\begin{equation}
\bar \mathcal{W}^\prime=\frac{1}{\pi
a^2}\int\limits_{-\infty}^{+\infty} \D^2\alpha\
P_\mathrm{out}(\alpha)\
\mathcal{W}\left(\alpha\right)=\frac{1}{\eta}P_\mathrm{in}
\left(\frac{\gamma}{\sqrt{\eta}},s^\prime\right),\label{UnbW}
\end{equation}
where
\begin{equation}
s^\prime=1-2\frac{\bar{n}(1-\eta)+a^2}{\eta}.
\end{equation}
This procedure can test the nonclassicality only if $s^\prime>-1$.
In other words, testing the nonclassicality with the witness
function (\ref{J}) is impossible if $a^2\geq
\eta-\bar{n}\left(1-\eta\right)$.

However, in case when the mean number of thermal photons $\bar{n}$
is less than the thermal threshold, one can apply the witness
function $\mathcal{W}_\mathrm{th}\left(\alpha\right)$ that is
given by Eq.~(\ref{K}). In an experiment, the value of
$\bar{\mathcal{W}}_\mathrm{th}$ can be reconstructed with using
Eq.~(\ref{Modified}). This gives a numerical result, which is
equal to Eq.~(\ref{FockBarW}), that indicates both the presence of
nonclassicality and non-positive values for the phase-space
distribution with $s=1-2\left(a^2\eta-\bar{n}(1-\eta)\right)$.

\section{Conclusions}
\label{S7}

Interaction of the system with an environment leads to the
disappearance of its nonclassical properties. Mostly this process
is known as decoherence, when off-diagonal matrix elements of the
density operator vanish in a certain representation. As a result,
initially pure quantum state evolves to a mixed one.

The phenomenon of nonclassicality has different manifestations
such as sub-Poissonian statistic of photon counts, photon
antibunching, quadrature squeezing, {\em etc}. Following to
\cite{Richter} we consider the state as nonclassical one if its
Glauber-Sudarshan $P$-function cannot be interpreted as
probability distribution. Indeed, for the specific quantum states
this function has negative values and, moreover, has very strong
singularities, which do not allow us to express it in terms of
regular functions.

If the environment has a non-zero temperature, the corresponding
system can be considered as being under the thermal noise
influence. In this case, the Glauber-Sudarshan $P$-function
smoothes, and domains of negative values as well as singularities
disappear. Eventually, beyond a certain value of the temperature
(thermal threshold), the system does not manifest any nonclassical
properties.

We consider the case of an oscillator-like system ({\em e.g.} a
mode of the electromagnetic field) interacting with hot
environment of other oscillators (other modes of field, absorbtion
system, {\em etc}). The Glauber-Sudarshan $P$-function of such a
system under the noise influence evolves according to the
diffusion-like equation, where the mean number of photons
$\bar{n}$ plays a role of ``the time variable'' as well as
``diffusion coefficient'' is expressed in terms of the
corresponding efficiency.

The most convenient method for testing the nonclassicality is the
measurement of a certain observable. Its normally-ordered symbol
is called  the witness func\-tion. Negative mean value of this
observable indicates nonclassical properties of the corresponding
quantum state. In principle, the concept of the witness function is
quite general and can be used for testing other forms of the
nonclassicality \cite{Korbicz}.

In case if the temperature is less than the thermal threshold and
noiseless state is tested for the nonclassicality with the witness
function $\mathcal{W}\left(\alpha\right)$, then there exits (but
not always) other witness function
$\mathcal{W}_\mathrm{th}\left(\alpha\right)$, which has the same
mean value for the noisy state as $\mathcal{W}\left(\alpha\right)$
for the noiseless one. This new witness function can be obtained
as a solution of a diffusion-like equation with negative diffusion
coefficient. This feature explains both the existence of the thermal
threshold of the nonclassicality and restrictions for the application
of the proposed method. Indeed, if the mean number of thermal
photons $\bar{n}$, that plays a role of ``time variable'' in this
equation, is greater than a certain value, then corresponding
solution may not be a positive-definite one and, moreover, has
strong singularities.

An example is the witness function chosen in a form of the Gauss
distribution with dispersion $a^2$. The corresponding mean, that
is a value of the phase space distribution with $s=1-2a^2$ in a
certain point, can be reconstructed in an experiment using
procedure of unbalanced homodyne detection \cite{Wallentowitz}.
The evolution of this witness function according to the diffusion-like
equation with negative diffusion coefficient results in decreasing
of the dispersion. It is clear that there exists a value of
$\bar{n}$ when it degenerates into the $\delta$-function. Beyond this value, the solution is defined in the space of distributions
which, moreover, are not positive-definite ones. Hence, testing
the nonclassicality beyond a certain threshold is impossible.

Another method for testing the nonclassicality is the application of
the discrete form of the Bochner criterion. This technique
requires to know values of the characteristic function in some
points. It can be realized via measurements of quadratures.
Corresponding procedures are developed for different systems
\cite{WelschRev}. Particularly, for the one-mode electromagnetic
field it can be performed with balanced homodyne detection.
However, in this case utilizing  the diffusion-like equation with
negative diffusion coefficient, one obtains the witness function
$\mathcal{W}_\mathrm{th}\left(\alpha\right)$, which is not
positive-definite one for any $\bar{n}>0$. This means that this
technique cannot be improved and the discrete Bochner criterion
should be applied directly.

\ackn D.Yu.V. and A.A.S. gratefully acknowledge Werner Vogel,
Lars M. Johansen and J\'anos Asb\'oth for the fruitful discussions. A.A.S. thanks the
President of Ukraine for the research stipend.

\end{document}